\documentclass[5p,twocolumn]{elsarticle}
\usepackage{amsmath}
\usepackage{graphicx} \graphicspath{{./}}
\usepackage{verbatim,listings}
\usepackage[usenames,dvipsnames]{xcolor}
\usepackage{hyperref}
\usepackage{xcolor}
\usepackage{enumitem}

\setlist[description]{leftmargin=\parindent,labelindent=\parindent}

\newcounter{bla}

\journal{Computational Materials Science}

\begin{document}

\begin{frontmatter}

\title{Full spectrum optical constant interface to the Materials Project}

\author[UW,SIMES]{J.~J.~Kas}
\author[UW,SIMES]{F.~D.~Vila}
\author[SIMES]{C.~D.~Pemmaraju}
\author[PNNL]{M.~P.~Prange}
\author[LBL]{K.~A.~Persson}
\author[LBL]{R.~X.~Yang}
\author[UW,SIMES]{J.~J.~Rehr\corref{cor1}}
\ead{jjr@uw.edu}
\cortext[cor1]{Corresponding author}

\address[UW]{Department of Physics, University of Washington, 
             Seattle, WA 98195, USA}
\address[SIMES]{Stanford Institute for Materials and Energies Sciences,
SLAC National Accelerator Laboratory, Menlo Park, CA 94025, USA}
\address[PNNL]{Physical and Computational Sciences Directorate, Pacific Northwest National Laboratory: Richland, WA 99352}
\address[LBL]{Materials Science Division, Lawrence Berkeley National Laboratory, Berkeley, CA 94720, USA}

\date{\today}

\begin{abstract}
Optical constants characterize the interaction of materials with light
and are important properties in material design.
Here we present a Python-based Corvus workflow for simulations 
of full spectrum optical constants from the UV-VIS to hard x-ray wavelengths
based on the real-space Green's function code FEFF10
and structural data from the Materials Project (MP).
The Corvus workflow manager and its associated tools provide an interface to FEFF10 and the MP database. The workflow parallelizes the FEFF computations of optical constants over all
absorption edges for each 
material in the MP database specified by a unique MP-ID. The workflow tools determine
the distribution of computational resources needed for that case.
Similarly, the optical constants for selected sets of materials can be computed
in a single-shot. To illustrate the approach, we present
results for nearly all elemental solids in the periodic table, 
as well as a sample compound, and compared with experimental results. 
As in x-ray absorption spectra, these results are interpreted in terms of an atomic-like
background and fine-structure contributions.

\end{abstract}

\begin{keyword}
Optical constants \sep Materials Project \sep FEFF \sep Corvus
\end{keyword}

\end{frontmatter}

\section{Introduction}

Optical constants characterize the frequency dependent interaction
between light and matter in the long-wavelength limit.
Thus they are often important characteristics in materials design.
These properties include the complex dielectric constant and 
index of refraction, as well as the energy-loss function, photoabsorption
coefficient, and optical reflectivity.\cite{hoc,adachi2012handbook,henke,chantler} Many other physical
properties can be derived from the optical constants, including
electron energy loss spectra (EELS), inelastic mean-free paths, the
atomic scattering amplitudes, and Hamaker constants for the van
der Waals interaction. For these purposes, tabulations of experimentally determined optical
constants are widely used.\cite{hoc,adachi2012handbook,henke,chantler} For practical reasons, however, such tabulations are limited to a relatively
small number of well characterized materials over limited spectral ranges
and limited environmental conditions. Thus here we focus on
broad spectrum theoretical treatments for
the systems defined in the Materials Project(MP) database.

Significant progress has been made in the fundamental theory
of optical properties
since the pioneering works of Nozieres and Pines,\cite{nozieres}
Adler,\cite{adler} and Wiser.\cite{wiser} In particular modern first-principles
calculations based on time-dependent density functional theory (TDDFT)
and the Bethe-SalpeterEquation (BSE) are now highly
quantitative.\cite{tddftvmbpt,MartinReiningCeperley,lawler,OCEAN,exc,exciting,octopus,RTSIESTA,AMBROSCHDRAXL20061,yambo}
Also limited tables of theoretical optical constants of
materials have been compiled over the optical
range.\cite{excitingTables,MaksimovTables}
However, such calculations involve many-body calculations of
optical response, and become computationally intractable
for many materials over broad spectral ranges.
In contrast, the real-space Green's function (RSGF) approach in
FEFF10 includes the key many-body effects, is highly automated and
has proved to be quite accurate for general materials in the x-ray regime.
 Moreover the code can be 
semi-quantitative in the UV-VIS to soft-x-ray regime.\cite{prange_opcons} 
Thus the code provides an efficient platform for 
calculations of dielectric properties over a broad spectral range.   
Consequently, theoretical calculations based on FEFF10 provide
an attractive alternative to available theoretical and experimental tabulations of optical constants for
many purposes. Nevertheless, selected experimental measurements are important
to validate the theory.

Recently K- and L-shell XAS calculations 
based on the FEFF code have   been added for a very large number of materials in the
Materials Project (MP) database.\cite{MP-K,MP-L} These data have been exploited,
e.g., in machine-learning models for the interpretation of XAS
data.\cite{MP-K,Torissi2020}  Our aim here is to complement these properties with a more complete set of optical constants for all edges from the UV-VIS to hard-x-ray energies.  Consequently, the results presented here provide
   significant extension, both in
the variety of optical-constant spectra,
spectral range, and material properties. Our approach is based on the the development of the Corvus \texttt{[opcons]} workflow, where Corvus is the workflow engine,\cite{Corvus} and \texttt{opcons} is the target property. In particular, the approach imports the structural and property data from the MP, and then parallelizes and automates the calculations permitting high-throughput calculations of optical constants for materials throughout the MP database.  
Our procedure for theoretical calculations of optical constants over all edges essentially follows that
described in detail by Prange \textit{et al.},\cite{prange_opcons} but has
been updated for the FEFF10 code.  In addition, refinements for including vibrational
disorder via the correlated Debye model using MP data have been added and the calculations for the UV-VIS range have been simplified. To illustrate
the approach calculations have been carried out for nearly all elemental
solids throughout the periodic table and explicit examples are given for
Cu, Ag, and Au, together with comparisons to experimental data. In addition we show
results for a sample compound Al$_2$O$_3$ to validate our
approximation in the UV-VIS range.
\begin{figure}[t]
  \includegraphics[scale=0.36,clip,trim=2.0cm 2.0cm 1.5cm 2.0cm]{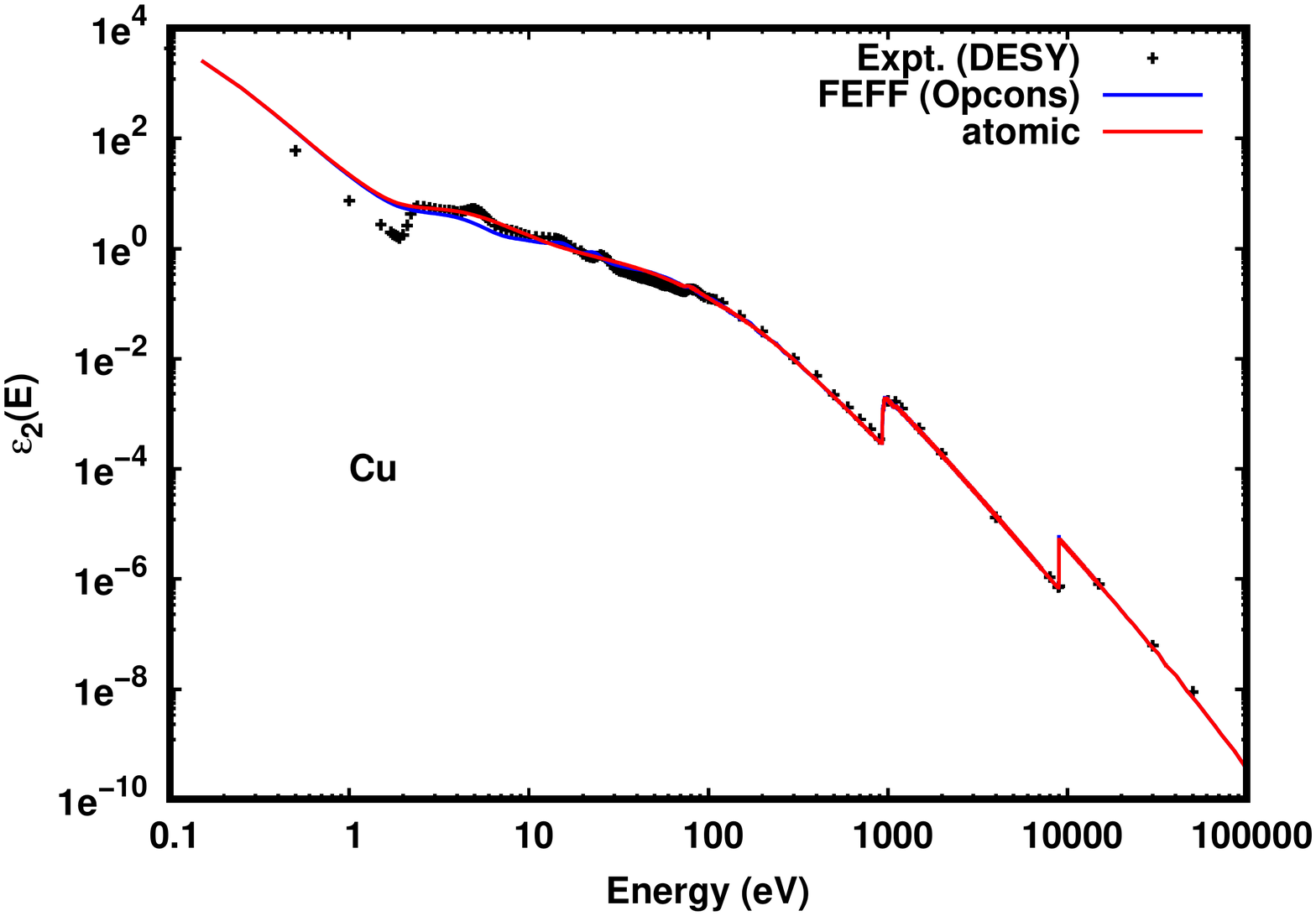}
  \includegraphics[scale=0.36,clip,trim=2.0cm 2.0cm 1.5cm 2.0cm]{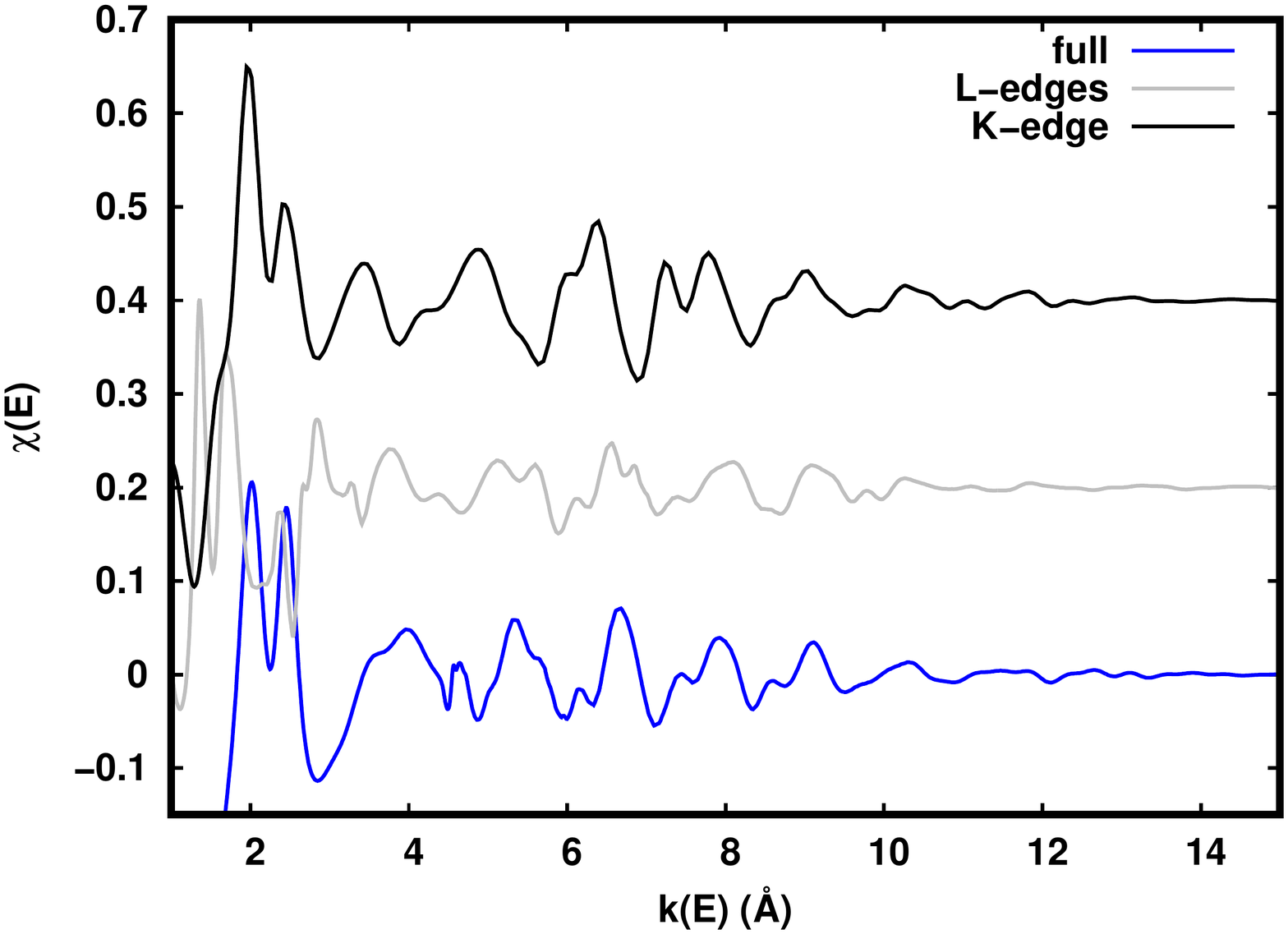}
\caption{Top: Calculated imaginary part of the dielectric function
  $\epsilon_2$ for Cu (blue) along with the atomic background $\chi$
(see text) (red). Bottom: Approximate optical fine structure of the
  imaginary (blue) part of $\epsilon$ for Cu vs $k=\sqrt{2E}$,
along with the same fine-structure defined with energies $E$ relative to the
K- and L-edges. Note the similarities between the fine structure found
in the optical and that of the K-edge. Note also that the
  fine structure becomes negligible for large $k$ above any
edges or outside the optical range. 
\label{fig:chi}
}
\end{figure}
The Corvus \texttt{[opcons]} workflow described here is applicable to
any material available in the MP database as defined by a given MP identificatoin label (MP-ID), thus making possible routine
calculations of optical constants for over $10^5$ structures. 
By default, the calculations are carried out with lattice vibration effects included
at room temperature, while electronic temperature and thermal expansion effects are neglected.
However, corrections for thermal expansion, finite temperature, and pressures other than 1 atm 
can be added using subsequent calculations that reuse some of the previously computed data.  The workflow is naturally parallelized and, for typical settings,
requires as little as minutes of wall-clock time per material on massively parallel
systems.


\section{Optical constants with Corvus and FEFF10}


In general, the optical constants are related to the complex
dielectric constant $\epsilon(\omega)=\epsilon_1(\omega)+i\epsilon_2(\omega)$
in the long wave-length limit,
where $\omega$ is the frequency of the electro-magnetic field. 
Since their real and imaginary parts are related by Kramers-Kronig (KK)
transformations, they can all be calculated from 
the imaginary part $\epsilon_2(\omega)$ (e.g.\ Fig.\ \ref{fig:chi}).
In this work the, the Corvus \texttt{[opcons]} workflow
yields the complex dielectric function $\epsilon(\omega)$, energy loss function $L(\omega)$, complex
index of refraction ${\bf n}(\omega)$,
 absorption
coefficient $\mu(\omega)$, and reflectivity $R(\omega)$ defined as
follows, in terms of $\epsilon_2(\omega)$, 
\begin{align}
  \label{eq:opcons}
  &\epsilon_1(\omega) = 1 +
  \frac{2}{\pi}{\cal P}\int_{0}^{\infty}d\omega'\ \frac{\omega'\epsilon_2(\omega')}{\omega^2-\omega'^2} \\
  &L(\omega) = -{\rm Im}[\epsilon^{-1}(\omega)] = \frac{\epsilon_2(\omega)}{\epsilon_1(\omega)^2+\epsilon_2(\omega)^2} \\
  &{\bf n}(\omega) = n(\omega)+i\kappa(\omega) = \epsilon(\omega)^{1/2} \\
  &\mu(\omega) = 2\frac{\omega}{c}\kappa(\omega) \\
  &R(\omega) = \frac{[n(\omega) - 1]^2+\kappa(\omega)^2}{[n(\omega) +1]^2+\kappa(\omega)^2} , 
\end{align}
where {$\cal P$} denotes the principal part of the integral.
Note that at high energies dielectric response is weak, and
$\epsilon_2(\omega) \approx L(\omega) \approx 2\kappa(\omega)
\approx \mu(\omega)(c/\omega)$.
 Sum-rules for the dielectric properties provide
a qualitative check on the reliability of the results.\cite{altarelli}
Related properties, such as the local atomic polarizability
$\alpha(\omega)=(\epsilon(\omega)-1)/(4\pi n)$, 
the Rayleigh forward scattering amplitudes
$f(\omega)= \omega\alpha(\omega)/r_0 c^2$ (where $r_0=e^2/mc^2$
is the classical
radius of the electron, $c$ is the speed of light and $n=N/V$ is the atomic number density), electron energy loss spectra (EELS),
inelastic mean-free paths, Hamaker constants $\epsilon(i\omega)$, etc.,
can also be determined, \cite{prange_opcons} but are not yet
implemented in the present Corvus \texttt{[opcons]} workflow. For simplicity here and below we use atomic units $e=\hbar=m=1$ throughout this paper unless otherwise needed for clarity.

The treatment of excitations from the core- and valence-levels
generally requires different considerations, so it is convenient to separate
the calculations as
\begin{equation}
\epsilon_2 = \epsilon_2^{\rm core} + \epsilon_2^{\rm val},
\end{equation}
where $\epsilon_2^{\rm core}$ includes the contribution from all levels below a fixed
core-valence separation energy $E_{cv}$, which in FEFF10
is set to -40 eV. Although this separation is approximate due to
many-body effects, the contributions from deep core levels (and to a reasonable
approximation, semi-core states) respond nearly independently from the
valence.

\subsection{Core level contributions}

In this paper calculations of the optical constants are carried out using
the real-space Green's function code FEFF10 and an updated version of
the approach described by Prange et al.\ \cite{prange_opcons}
FEFF uses the same real-space Green's function
formalism to calculate a variety of spectroscopies and related
quantities including XAS, XES, RIXS, EELS, and Compton. Thus the
calculations of the optical constants use essentially the
same ingredients as those for the well-established calculations of XAS,
greatly simplifying the workflow design.

Briefly, the RSGF approach in FEFF is used to calculate the
contribution to $\epsilon_2(\omega)$ from each occupied level $i$
(i.e., absorption edge) at each unique site in the system $a$.
For simplicity, the site index will be usually suppressed below.
The core level contribution to $\epsilon_2(\omega)$ from a given site $a$
consists of a sum of contributions from occupied core levels
$|i\rangle$ \cite{prange_opcons} for $\varepsilon_i < E_{cv}$
given by
\begin{equation}
\label{eq:eps2core}
  \epsilon_{2}^{\rm core}=\sum_i\frac{4\pi}{\omega}{\rm Im}\left[\langle i|{\hat d}^\dagger G(\omega
    + \varepsilon_i){\hat d}|i\rangle\right]\theta(\omega - \varepsilon_F+\varepsilon_i),
\end{equation}
where $|i\rangle$ is the core-orbital of interest,
$\varepsilon_i$ its binding energy, $\hat d$ is the
transition operator, which in most cases is dipole dominated, and
$G(\omega)$ is the one-electron Green's function. Finally, 
$\theta(\omega-\varepsilon_F+\varepsilon_i)$ is a unit step function
that turns on when the photon energy $\omega$ is sufficient to 
excite the core-electron above the Fermi energy $\varepsilon_F$. 
If the interaction between the core- and photoelectron levels is ignored,
this level of approximation is equivalent to the RPA. However, in this
work we use the final-state rule
approximation to the core hole interaction, in which the photoelectron Green's function is calculated
in the presence of a self-consistently screened core-hole.

One of the advantages of the RSGF approach is the separation of the propagator  
$G(\omega)$ into contributions from the central absorber at a given 
site and single or multiple-scattering contributions from neighboring
atoms, i.e., $G(\omega) = G^{\rm abs}(\omega) + G^{\rm sc}(\omega)$. As a result
the dielectric properties exhibit fine structure analogous to that in
XAFS, i.e.,
\begin{equation}
\epsilon_2 = \sum_i \epsilon_{2i}^{\rm atomic} [1 + \chi_i(\omega)],
\label{eq:opticalfs}
\end{equation}
which we dub ``optical fine structure" (OFS).
Typically $\chi_i(\omega)$ is of order a few percent within a 
range of order 100 eV above each edge; however, $\chi_i(\omega)$ can be
large and of order unity in the near-edge regime, within about 10 eV
or so of a given edge.  Thus the physical interpretation of the OFS is similar
to that for EXAFS and XANES, reflecting the local geometrical structure in the
vicinity of an absorbing atom.  These observations also explain why a
local, atomic approximation that neglects the fine structure
from neighbor scattering
can be a reasonable approximation, accurate to within a few percent
for the core contributions to the optical constants, except close to
an absorption edge.  In order to show the fine
structure over the entire energy range of our calculations, it is
convenient to calculate the fine-structure defined as $\chi_2 =
(\epsilon_2 - \epsilon_2^{\rm atomic})/\epsilon_2^{\rm atomic}$,
which is shown for Cu in Fig.~\ref{fig:chi} (bottom), together with
the full and atomic approximation to $\epsilon_2$ (top). 
A corresponding approximation can also be made
for any of the other optical constants.  When
$\epsilon_1^{\rm core} \approx 1$, the fine structure in the loss function
$L(\omega)$, absorption $\mu$ and imaginary part of the index of
refraction $\kappa$ are similar to that in $\epsilon_2$. The fine
structure in $\epsilon_1$ is similar, although phase shifted by
90$^{\circ}$. 

\begin{figure}[t]
\includegraphics[scale=0.36,clip,trim=2.0cm 2.0cm 1.5cm 2.0cm]{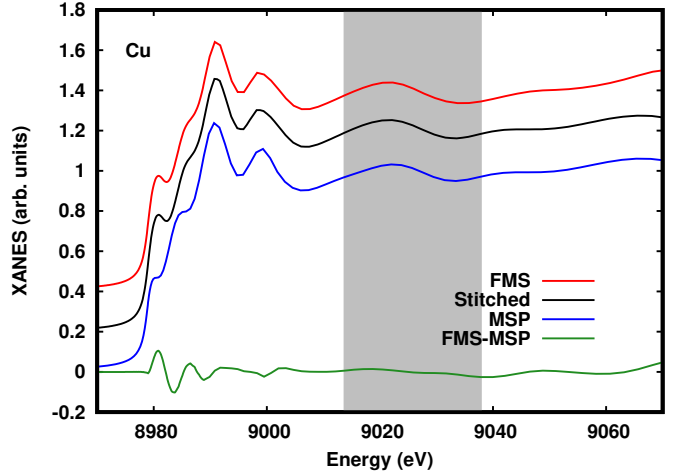}
\caption{Cu K-edge spectrum
  obtained using the stitching algorithm (black). For comparison,
  the curves calculated using FMS (red) and MSP (blue) are also shown,
  shifted above and below the stitched curve, respectively. The green
  curve is the difference between the FMS and MSP calculated results, and
  the shaded area shows the region where the stitching algorithm is applied.
\label{fig:CuK}
}
\end{figure}

In order to obtain optical constants over a broad energy range from
the UV to hard x-rays, two FEFF calculations must be performed for each
edge (\textit{i.e.} each occupied atomic level $i$) and each unique atomic site $a$ in the
presence of the corresponding core-hole: 1) In the near edge regime
within about 50 eV of a given edge full-multiple-scattering
(FMS) calculations must be performed, and 2) at higher energies
(the EXAFS regime), the multiple-scattering path expansion (MSP) is
used. Once these calculations are performed, the results are
interpolated onto a common
energy grid and stitched together to yield the spectrum
$\epsilon_{2}^i(\omega)$ coming from the core-state $|i\rangle$.
The stitching algorithm used is:\cite{prange_opcons} 
\begin{align}
  \epsilon^{i}_2(k) =& \epsilon_{2}^{{\rm FMS},i}(k);
&k<k_0, \nonumber \\
=& \cos^2(\frac{k-k_0}{k_1-k_0})\epsilon_{2}^{{\rm FMS},i}(k) \nonumber \\
  &+\sin^2(\frac{k-k_0}{k_1-k_0})\epsilon_{2}^{{\rm MSP},i}(k);
& k_0 \le k \le k_1, \nonumber \\
=& \epsilon_{2}^{{\rm MSP},i}(k); &k>k_1
\end{align}
where $k_0= 3~{\rm \AA}^{-1}$,
$k_1 = 4~{\rm \AA}^{-1}$,
and $k=[2(E-E_{\rm edge})]^{1/2}$ is the photoelectron wave-number.
This is illustrated in Fig.\ \ref{fig:CuK} for the K-edge spectrum of fcc Cu.

\begin{figure}[t]
\includegraphics[scale=0.35,clip,trim=1.8cm 1.2cm 1.0cm 2.3cm]{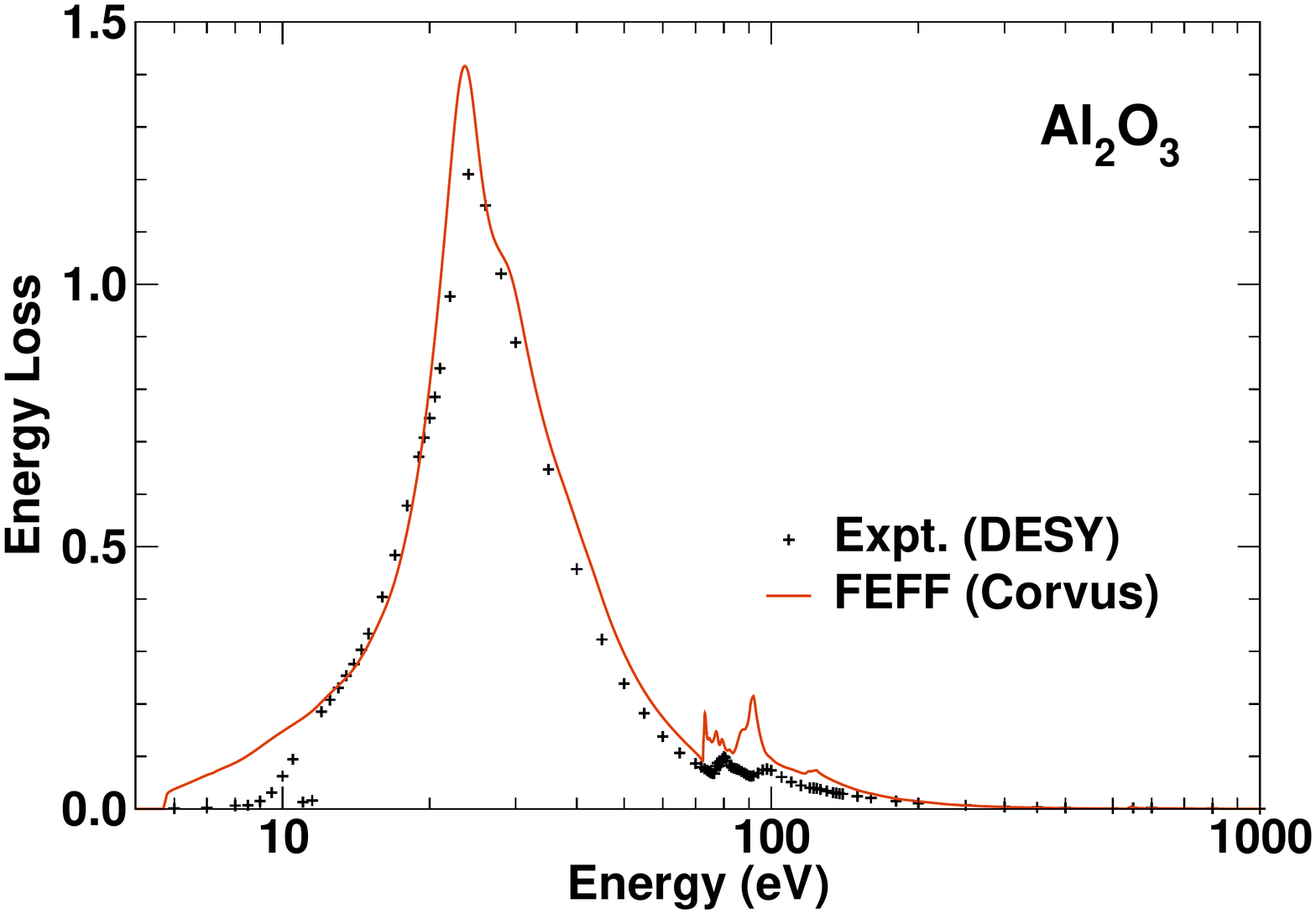}
\includegraphics[scale=0.35,clip,trim=1.8cm 1.2cm 1.0cm 2.3cm]{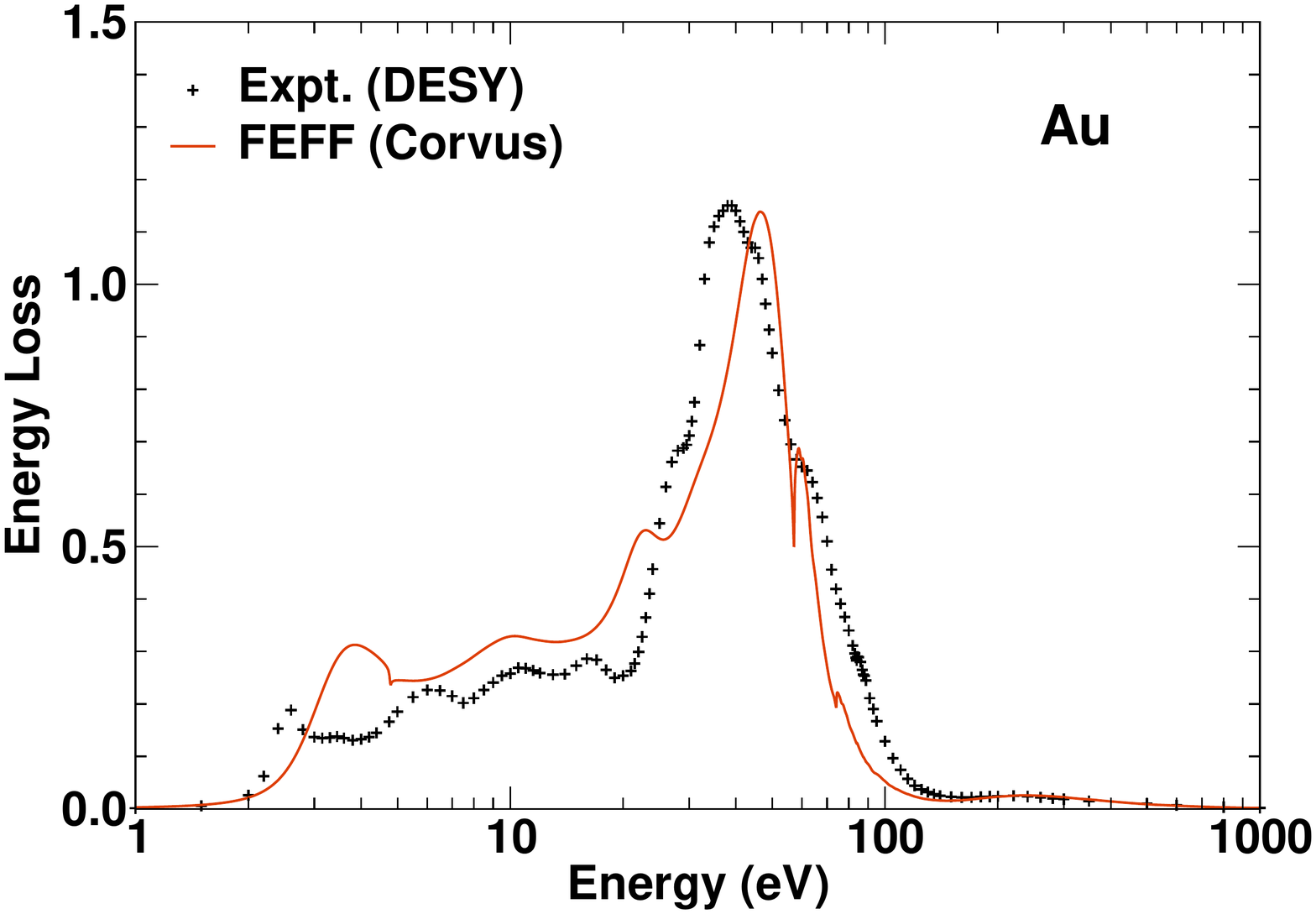}
\caption{Comparison of the experimental\cite{desy} and calculated
energy loss function  for Al$_2$O$_3$ (corundum) (top)   and Au (gold) (bottom)  with the
local approximation (see text).}
\label{fig:Al2O3loss}
\end{figure}

\subsection{Valence Contributions}

The contribution from the valence levels is complicated by 
the extended range and continuum nature of the valence wave functions
and the large density of states in the valence region.
In the full RSGF approach this requires sums over multiple sites as in
Eq.\ (32) of Ref.\ \cite{prange_opcons}. Moreover, systems with strong
excitonic effects often require treatments beyond the RPA such as TDDFT 
or the Bethe-Salpeter equation.\cite{MartinReiningCeperley} Several codes are available for these advanced treatments,\cite{lawler,OCEAN,exciting,RTSIESTA} 
but they are computationally challenging for high throughput calculations.
Nevertheless, neglecting these effects is often a reasonable approximation
for semi-quantitative purposes. Consequently, in this work we
use a local approximation
at the absorption site to calculate the contribution from the valence
electrons. In addition, we replace the valence wavefunctions with the
Dirac-Fock atomic eigenfunctions, which allows us to write
Eq.\ (32) of Ref.\ \cite{prange_opcons} 
as an integral over the local valence density of
states \begin{equation}
\label{eq:dosconv}
\epsilon_2^{\rm val}(\omega) = \sum_{ia}\int_{E_{cv}}^{E_F} d\omega' \tilde\rho_{ai}(\omega')
\epsilon_2^{ia}(\omega - \Delta - \omega'),
\end{equation}
where $\epsilon_2^{ia}(\omega)$ is the local atomic spectrum
at a given site $a$ corresponding to atomic state $i$, and
$\tilde\rho = \rho/N$ is the normalized,
angular momentum projected density of states (LDOS) with symmetry
equal to that of the state $i$, and
$\Delta = E_{\rm top}-E_{\rm edge}$ is a shift
to properly position the valence spectra, where $E_{\rm top}$ is the
energy at the top of the gap, if one exists, and is equal to the Fermi
energy otherwise.
\begin{figure}[t]
  \includegraphics[width=\columnwidth,clip,trim=1.8cm 2.0cm 1.0cm 2.1cm]{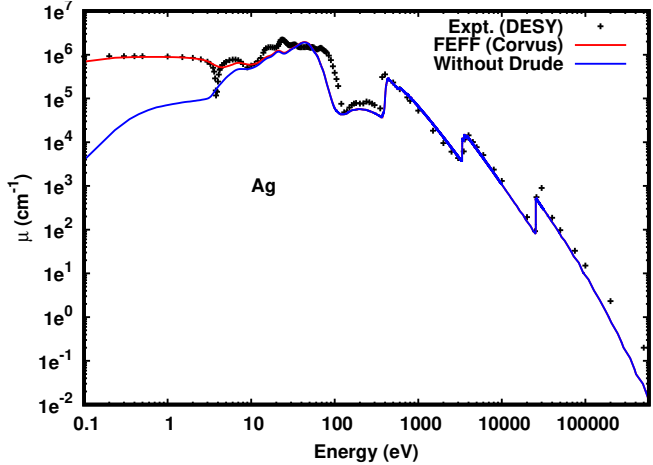}
  \caption
      {Comparison of the theoretical absorption $\mu$ of Ag with and without the Drude contribution, along with experimental results \cite{desy}.}
      \label{fig:Agmu}
\end{figure}
As shown for the energy loss of Al$_2$O$_3$ (top) and Au (bottom) in Fig. \ref{fig:Al2O3loss}, which provide good tests for this procedure we find that the approximation is
reasonably accurate when excitonic effects are small. We have found that
our local approximation in Eq.\ (\ref{eq:dosconv}) is significantly better than a pure atomic approximation alone.
One can also define an optical fine structure $\chi^{\rm val}$ for
the valence contributions $\epsilon_2^{\rm val}$, although its magnitude
is typically large and of order unity (see Fig.~\ref{fig:chi}), as in core-level XANES, over
the optical energy range.  For this reason, FMS and the
local projected densities of states are needed to
evaluate Eq.\ (\ref{eq:dosconv}). In addition, the physical interpretation
of $\epsilon_2^{\rm val}$ is akin to a sum of terms proportional to the
joint projected density of states.\cite{prange_opcons} Even so, the valence fine
structure for Cu is rather similar to the K-edge fine structure, as
shown in the lower panel of Fig.~\ref{fig:chi}. 
This makes sense, since the $d$- to $p$-state
transitions dominate in the optical region, and the DOS is dominated
by a single, relatively sharp occupied $d$-band. 

Finally, the total $\epsilon_2(\omega)$ is obtained from 
a double sum over edges $i$ and unique sites in the system,
\begin{equation}
  \epsilon_2(\omega) = \sum_{ia} n_a \epsilon_2^{ia}(\omega), 
\end{equation}
where $a$ denotes a unique site in the unit cell, and $n_a=N_a/V$
is the corresponding number density.

\begin{figure}[t]
\includegraphics[scale=0.35,clip,trim=1.8cm 1.2cm 1.0cm 2.3cm]{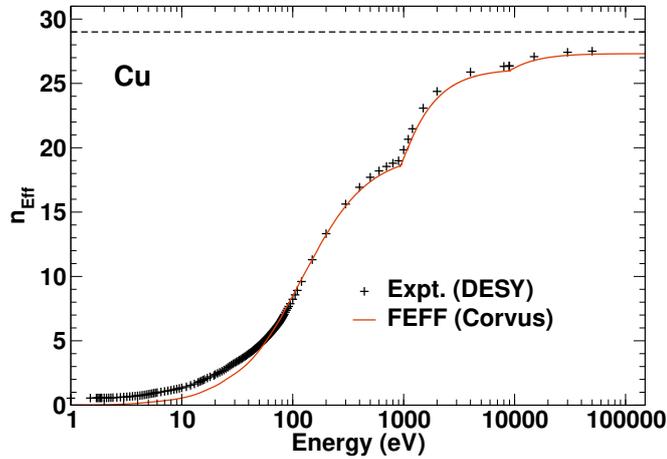}
\caption[Sum rule for Cu]
{Comparison of the experimental\cite{desy} and theoretical $\epsilon_2(\omega)$ sum rule $n_{\rm eff}(\omega) =\int_0^\omega d\omega\, \epsilon_2(\omega)/\pi$ 
for Cu. The horizontal dashed line indicates the ideal theoretical limit ($Z=29$).}
\label{fig:Cusum}
\end{figure}

\subsection{Debye-Waller factors and Drude corrections}

A number of other considerations are generally important in calculations of
optical spectra over broad energy ranges. For example,
effects of thermal vibrations are temperature dependent and
strongly damp the fine-structure in the EXAFS regime.
For the systems treated here
we have approximated the multiple-scattering path dependent
DW factors $\exp(-2\sigma^2 k^2)$
using the default correlated Debye model
in FEFF10. The Debye temperature parameter can be
estimated\cite{Debye-tempW-moduli} from the bulk and shear
elastic moduli data available in the MP database.

Drude corrections in metals can be added phenomenologically
by an additional {\it ad hoc} contribution to the limiting low
frequency behavior of $\epsilon_2(\omega)$ given by
\begin{equation}
\epsilon_2^{D}= \frac{4\pi\sigma}{\omega}= \frac{\omega_p^2 \tau}{\omega}
\end{equation}
 where $\sigma$ is the conductivity, $\omega_p^2 =4\pi n e^2/m$,
$n$ is the mean conduction electron density, and $\tau$ the relaxation
 time, which is typically of order $0.1-4.0$ femtosecond
at room temperature.\cite{ashmem} Fig.~\ref{fig:Agmu} shows a comparison of the absorption $\mu$ of Ag with and without the Drude contribution, along with experiment.\cite{desy}

Thermal expansion can also be treated by calculating the expansion
parameter via DFT,\cite{dmdw} or using parameters from the literature. 
Additional broadening due to
electron-phonon
interactions, impurities, and other mechanisms can also be important.
While neglected by default, these effects can be added {\it ex post facto}
by a Lorentzian broadening of the spectra.

Once $\epsilon_2(\omega)$ is obtained by summing both core and valence
contributions over all sites, the real part of the dielectric
function is formed via a KK-transform, as in Eq.~(\ref{eq:opcons}). 
A check on the completeness of the calculations is provided by
the optical constant sum-rules,\cite{altarelli} as illustrated in
Fig.\ \ref{fig:Cusum}.

\begin{figure}[t]
\includegraphics[width=1.0\columnwidth]{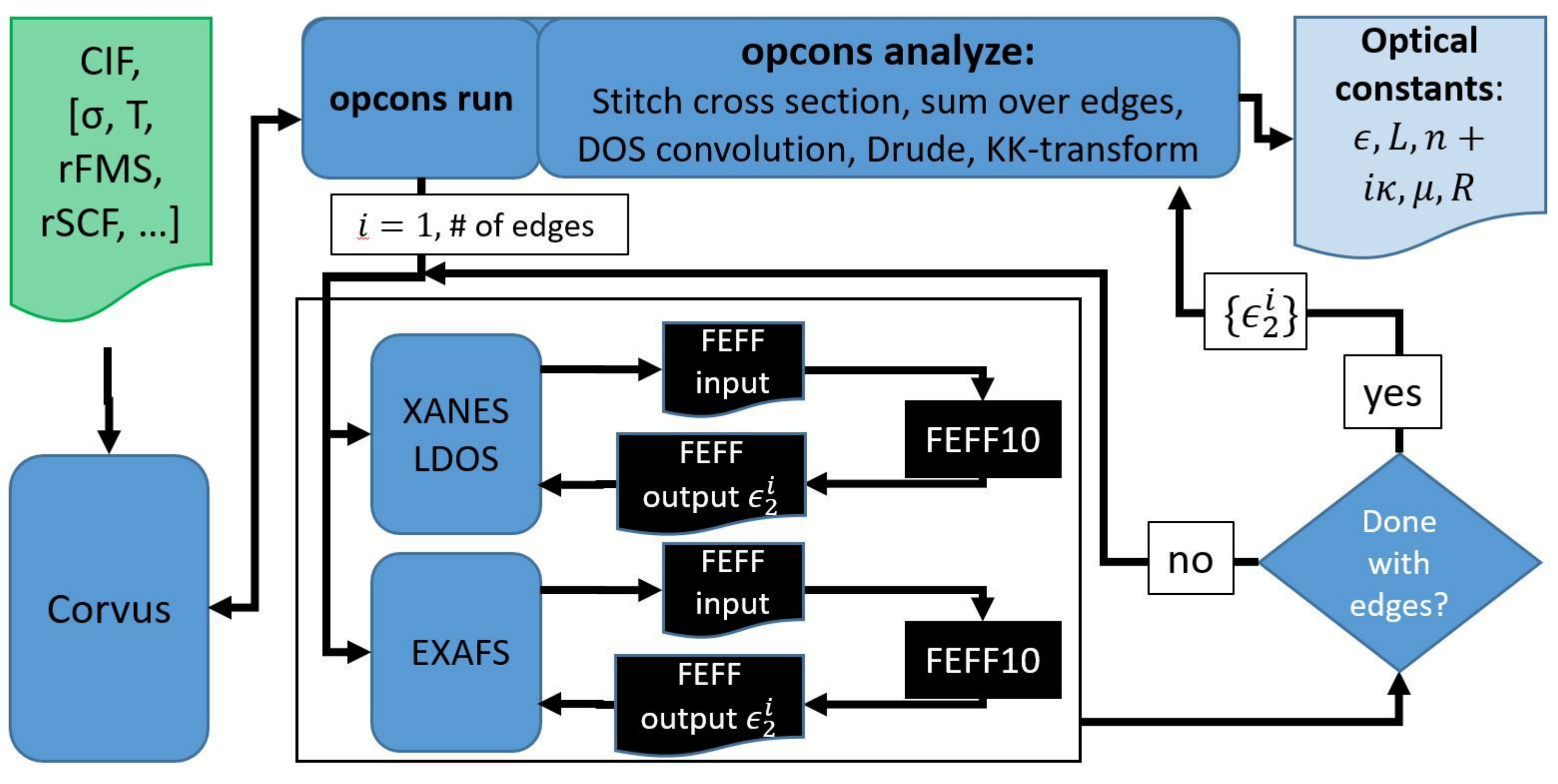}
\caption
{Schematic of the Corvus workflow for optical constants. User input is
  shown in light green, while code blocks associated with Corvus are
  shown in blue, and input/output/code associated with FEFF is shown
  in black. The output is shown in light blue. Note that Corvus \texttt{[opcons]}
  requires minimal input from the user (only the CIF file and a
  request for optical constants), although many optional
  parameters are available.
}
\label{fig:opcons_wf}
\end{figure}

\subsection{ {\rm Corvus  \texttt{[opcons]}} workflow } 

The Corvus \texttt{[opcons]} workflow is built
on the concept of recursive workflow generation, wherein Corvus generates
and runs a workflow with the target property \texttt{opcons}, and
the optical constants   workflow \texttt{[opcons]} in turn
generates and runs a \texttt{[xanes]} workflow for each required FEFF10 calculation. 
The workflow tool is described in detail together with various examples in the Corvus reference.\cite{Corvus}
This approach determines the input files from the data in the MP database, afterwhich all of the individual FEFF10 calculations are set up and run
through the Corvus workflow manager. The Corvus \texttt{[opcons]} workflow then carries out the following steps: i) calculating the $\epsilon_2(\omega)$ for a given edge; ii) applying the 
stitching algorithm to combine the near-edge full multiple scattering (FMS) and multiple-scattering path (MSP) calculations;
iii) convolving with the density of states (Eq.~\ref{eq:dosconv}); iv) performing 
the KK-transform to obtain $\epsilon_1$; v) adding the Drude
contribution; and vi) calculating the
other optical constants following the relations
in Eqs.\ (\ref{eq:opcons}). In order to facilitate automated
calculations, Corvus handles the creation of FEFF10 input
files from minimal user input, executes FEFF10, and translates the
output. Additionally, the Corvus
\texttt{[opcons]} workflow is designed to optimize the calculations, for example
re-using output from the XANES results within the EXAFS
calculations to bypass the self-consistency loop.
Figure~\ref{fig:opcons_wf} shows a schematic of
the Corvus \texttt{[opcons]} workflow for producing the optical constants.

%
%



\begin{figure}[t]
\begin{lstlisting}[language=sh,%
		   frame=single,%
		   basicstyle=\linespread{1.2}\scriptsize\ttfamily,%
		   commentstyle=\color{RoyalBlue}%
		  ]
Query found 16 systems
Keeping 3 systems after energy screening
Found systems:
# -------------------------------- #
System:              mp-1279742
Reduced Form.:       FeO
Cell Comp.:          Fe4 O4
Sym. Unique Comp.:   Fe1 O1
Total num. of edges:  15
Estimated Debye T:    466.60 (Pred.)
# -------------------------------- #
System:              mp-715262
Reduced Form.:       FeO
Cell Comp.:          Fe2 O2
Sym. Unique Comp.:   Fe1 O1
Total num. of edges:  15
Estimated Debye T:    416.52 (Calc.)
# -------------------------------- #
System:              mp-1178232
Reduced Form.:       FeO
Cell Comp.:          Fe4 O4
Sym. Unique Comp.:   Fe1 O1
Total num. of edges:  15
Estimated Debye T:    465.00 (Pred.)
# -------------------------------- #

Keeping 3 systems after symmetry and edge screening

Total number of edges to compute in set: 45
\end{lstlisting}
\caption{Typical output for the \texttt{crv\_mp\_mk\_set} command showing that the MP query returned 16 possible systems, of which only 3 are left after screening for energy, symmetry, and total number of edges. Also shown are the estimated Debye temperatures, which can be calculated (e.g., from DFT data), or predicted (e.g., from machine learning data).}
\label{fig:feoout}
\end{figure}

\section{Materials Project interface tools}

\subsection{Global structure of {\rm Corvus [opcons] } simulations}

MP optical constants simulations are set up in two steps: 1) The \texttt{crv\_mp\_mk\_set} command reads the command line inputs, creates a request to the MP, downloads all the structures available matching the query, generates a set of materials to run and calculates the totl number of edges in the set. 2) The \texttt{crv\_mp\_run\_set} command runs all the calculations required for the set, distributed in serial or parallel mode according to the resources provided. In the following section we describe their usage in detail.

\subsection{Usage: crv\_mp\_mk\_set}

The basic input of a typical Corvus \texttt{[opcons]} calculation is set up using the \texttt{crv\_mp\_mk\_set} command:

\texttt{crv\_mp\_mk\_set [OPTIONS] Set\_Name}

\noindent where ``\texttt{Set\_Name}'' is a text label that will be used throughout to refer to the set of materials. This label is used to name the directory where the calculations will be done and the results stored. \texttt{crv\_mp\_mk\_set} can be used to query the MP database in three different ways (see options below): 1) Query an n arbitrary class of materials selected by formula, 2) automatically query some or all of the elemental solids, for which the tools have an internal database of MP-IDs, and finally, 3) an individual material with a given MP-ID.
The full set of options can be obtained with ``\texttt{crv\_mp\_mk\_set --h}''. Here we focus on some of the most computationally relevant ones.

\begin{figure}[t]
\begin{lstlisting}[language=sh,%
		   frame=single,%
		   basicstyle=\linespread{1.2}\scriptsize\ttfamily,%
		   commentstyle=\color{RoyalBlue}%
		  ]
target_list { opcons }
usehandlers { Feff }
opcons.usesaved{ True }
cif_input{CIF_symm.cif}
feff.scf{ 3.2350952042425325 0 30 0.1 0 }
feff.fms{ 5.075798356336175 0 0 0.0 0.0 40.0 }
feff.debye{ 298.0 416.52459701743817 0 }

# The lines below should set the parallel settings
...
\end{lstlisting}
\caption{Typical template Corvus \texttt{[opcons]} input generated by the \texttt{crv\_mp\_mk\_set} command. This input shows the selected Corvus target \texttt{opcons}, the handlers that sould be used (only FEFF is required for these types of runs), the location of the structure in CIF format, and the FEFF SCF and FMS radii. The latter are automatically generated based on the number of shells requests in the command line. Depending on the options used, the calculation can use a Correlated Debye model to add thermal broadening, using a Debye temperature parameter estimated from MP database values of the bulk and shear moduli.
Given that Corvus stores the MPI command, number of processors, etc in the input, the actual input is generated at runtime by the \texttt{crv\_mp\_run\_set}, where the required parameters are added at the end of the template input shown here.}
\label{fig:opcin}
\end{figure}

\subsubsection{Required options}

\begin{description} \itemsep5pt \parskip0pt \parsep0pt
\item[\texttt{--k APIKEY}] \hfill \\
  API key used to interact with the MP REST interface
\end{description}
This option is required for the command to be able to interact with the MP database. See section \ref{subsubsec:req} for details. 

\subsubsection{System selection options}

\begin{description} \itemsep5pt \parskip0pt \parsep0pt
\item[\texttt{--f FORMULA}] \hfill \\
  Use chemical formula \texttt{FORMULA} to define a set of materials.
 \item[\texttt{--pt}] \hfill \\
  Create a set for all the solid elements in the periodic table using the structure for the most stable phase in standard conditions
\item[\texttt{--ptel PT\_ELEMS}] \hfill \\
  Same as \texttt{--pt}, but only for a list of specific elements
\item[\texttt{--mpid MP-ID}] \hfill \\
  Select the individual system with Materials Project identification label MP-ID (Default: ``mp-30'')
\end{description}
These options are mutually exclusive. If none of them are present the command generates input for a test run on Cu, a traditional choice since FEFF was first deployed.
 
\subsubsection{Selection fine-tuning options}

\begin{description} \itemsep5pt \parskip0pt \parsep0pt
\item[\texttt{--mxpt MXNAPT}] \hfill \\
  Maximum atomic number to use while creating the periodic table with \texttt{--pt} (Default: 99)
\item[\texttt{--enpct EN\_PCT}] \hfill \\
  Keep only \texttt{EN\_PCT}\% of lowest energy systems (Default: 100\%)
\item[\texttt{--nedg MXNEDGES}] \hfill \\
  Keep only systems with fewer than \texttt{MXNEDGES} edges (Default: 250)
\end{description}
These options help trim the raw selection generated by the system selection options.

\subsubsection{Auxiliary options}

\begin{description} \itemsep5pt \parskip0pt \parsep0pt
\item[\texttt{--nSCF NSCF}] \hfill \\
  Number of neighbor shells to use in SCF (Default: 1)
\item[\texttt{--nFMS NFMS}] \hfill \\
  Number of neighbor shells to use in FMS (Default: 2)
\item[\texttt{--symprec SYM\_PREC}] \hfill \\
  Symmetry precision (in \AA) used by symmetrizer (Default: 0.01\AA)
\item[\texttt{--dw}] \hfill \\
  Use the correlated Debye thermal broadening model in the optical constants calculation
\item[\texttt{--temp TEMPERATURE}] \hfill \\
  Temperature (in K) used for correlated Debye thermal broadening (Default: 298K)
\end{description}
These options control the setup of the optical constants calculations by defining how many shells to use in the SCF and FMS steps, as well as the precision of the symmetrization procedure used to reduce the number of unique sites. The shell selection is performed using the Jenks natural breaks classification method\cite{jenks1967data} as implemented in the \texttt{jenkspy} Python module.\cite{jenkspy} Additionally, vibrational broadening can also be controlled by requesting that a correlated Debye model\cite{sevillano1979extended} be used.

\begin{figure}[t]
\includegraphics[scale=0.32,clip,trim=0.6cm 0.0cm 0.0cm 0.0cm]{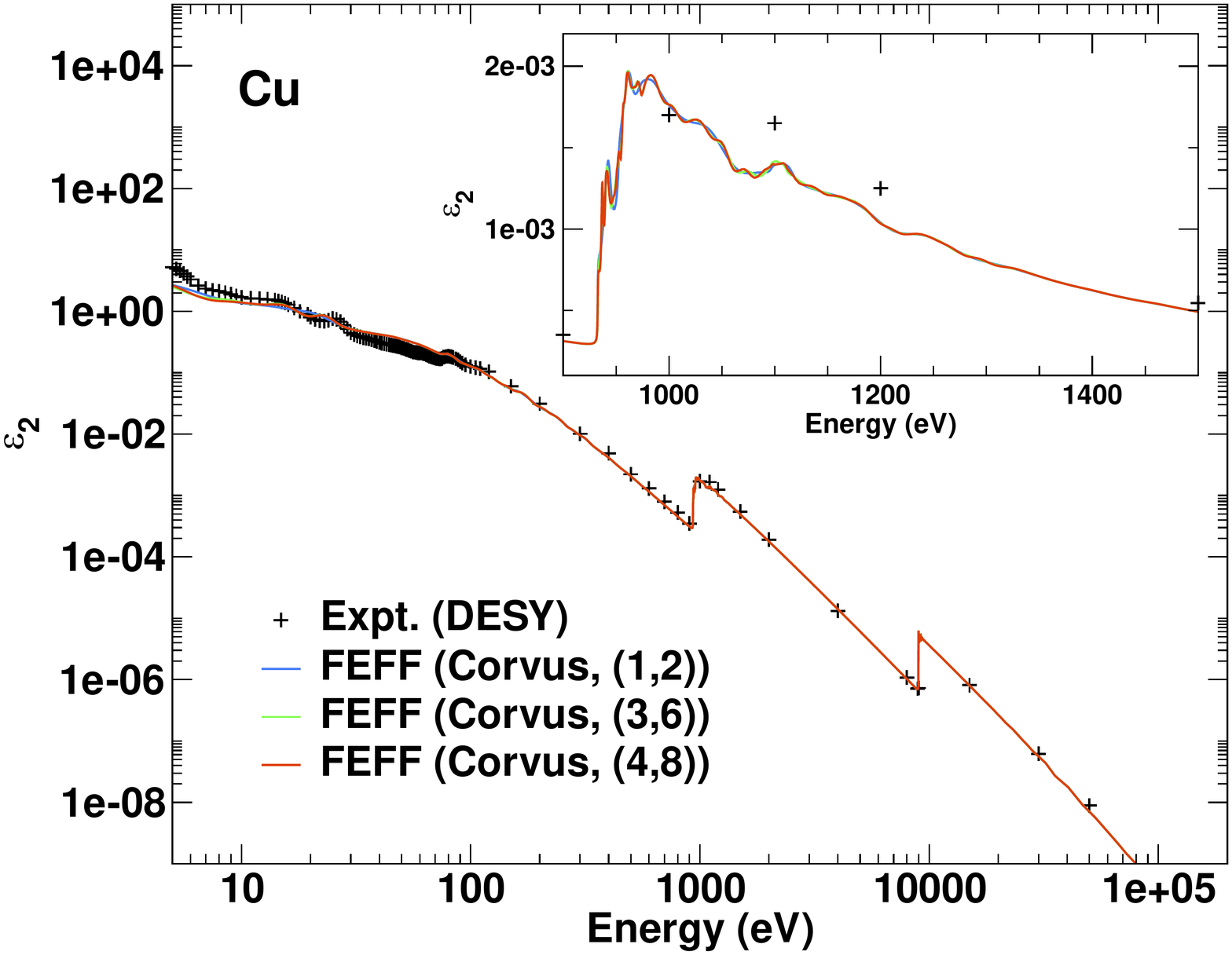}
\includegraphics[scale=0.36,clip,trim=1.8cm 1.2cm 1.0cm 2.5cm]{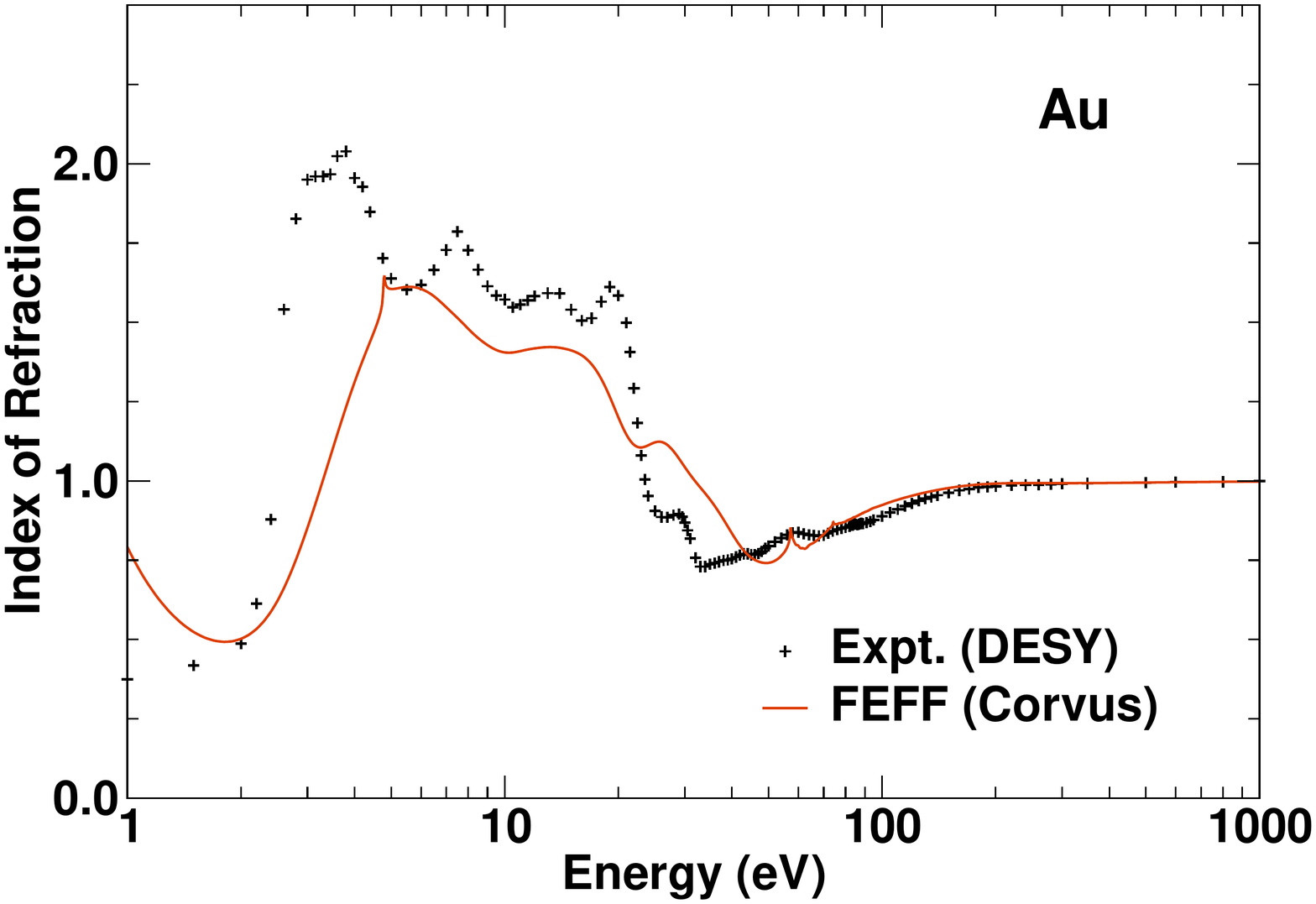}
\caption[$\epsilon_2$ for Cu]
{Comparison of the experimental\cite{desy} and theoretical $\epsilon_2$ for Cu as a function of FEFF Corvus (nSCF,nFMS) shells to explore convergence (top)
and (bottom) index of refraction for Au in the valence region (bottom).
}
\label{fig:rconv}
\end{figure}

\subsubsection{Calculation control options}

\begin{description} \itemsep5pt \parskip0pt \parsep0pt
\item[\texttt{--dr}] \hfill \\
  Dry run, collect and display information without actually creating a set
\item[\texttt{--v}] \hfill \\
  Make the command output more verbose
\end{description}

\begin{figure*}[t]
\includegraphics[scale=0.80,clip]{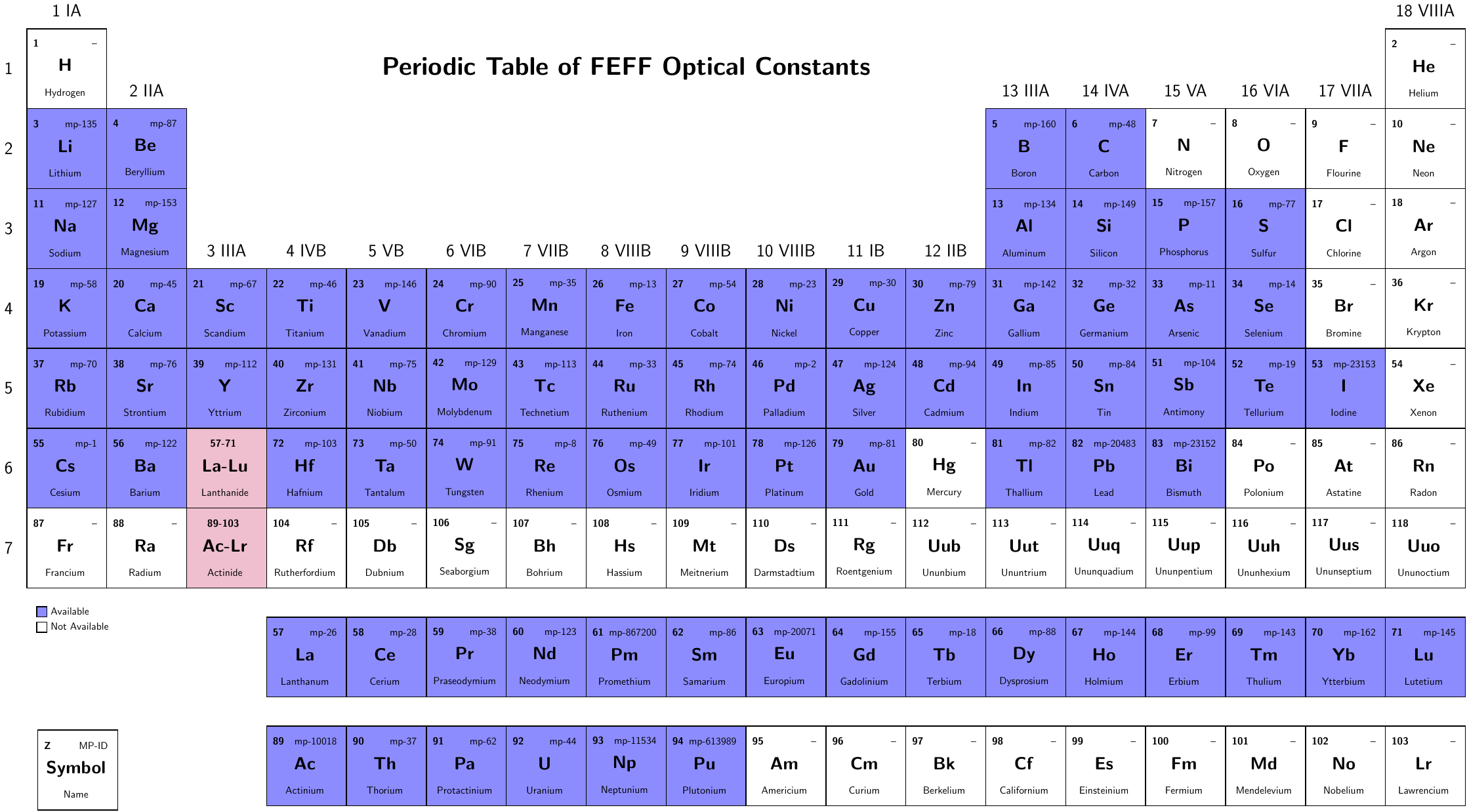}
\caption[Periodic Table of FEFF Optical Constants]
{Periodic table of FEFF optical constants currently available.
The index on the top-right corner of each element indicates the
MP-ID of the structure used in the calculations. A dynamic version of this table can be found at \url{http://feff.phys.washington.edu/optical-constants/},\cite{opconsurl} including all optical constants data as well as the input files used to generate it.}
\label{fig:ptable}
\end{figure*}

\subsubsection{Examples}
A typical invocation of \texttt{crv\_mp\_mk\_set} is as follows

\texttt{crv\_mp\_mk\_set --k [APIKEY] --f "FeO" --enpct 20 --nedg 250 --nSCF 2 --nFMS 4 --v --dw --temp 298.0 IronOxides}

\noindent creates a set of 3 systems (Figure \ref{fig:feoout}) after screening for energy, symmetry and total number of edges. The command also requests that the Debye temperature be estimated, and that the calculation be done at room temperature. Inside the set directory, the structures are organized by reduced formula and MP-ID. A typical Corvus \texttt{[opcons]} directory created with \texttt{crv\_mp\_mk\_set} contains, for each material, a template Corvus \texttt{[opcons]} input (Fig.\ \ref{fig:opcin}) and a few auxiliary files with structural information. A set for the Li, Be and B elemental solids can be generated using:

\texttt{crv\_mp\_mk\_set --k [APIKEY] --v --ptel 'Li,Be,B' LiBeB}

\noindent while a single compound (one of the forms of Mg$_3$N, in this case) can be set up with:

\texttt{crv\_mp\_mk\_set --k [APIKEY] --v --mpid mp-1185783}

\noindent More examples can be found in Ref. \cite{opconsinstall}.

\subsection{Usage: crv\_mp\_run\_set}

After a set is created, the actual optical constants can be computed with the \texttt{crv\_mp\_run\_set} command:

\texttt{crv\_mp\_run\_set [OPTIONS] Set\_Name}

\noindent where ``\texttt{Set\_Name}'' is again the label for the set to run.
As for ``\texttt{crv\_mp\_mk\_set}'', the full set of options can be obtained with ``\texttt{crv\_mp\_run\_set --h}'', but here we focus on some of the most computationally relevant ones.

\subsubsection{Options to control the number of processors}

\begin{description} \itemsep5pt \parskip0pt \parsep0pt
\item[\texttt{--np NP\_TOT}] \hfill \\
  Total number of processors to be used in the run (Default: 1)
\item[\texttt{--ppn PPN}] \hfill \\
  Number of processors per node (Default: 1)
\item[\texttt{--nn NNODES}] \hfill \\
  Number of nodes with \texttt{PPN} processors per node to be used in the run (Default: 1)
\end{description}
If more than one of these options are specified,
the command will choose \texttt{max(NP\_TOT,PPN*NNODES)} as the total number of processors to use.

\subsubsection{Options to estimate computational cost}

\begin{description} \itemsep5pt \parskip0pt \parsep0pt
\item[\texttt{--t}] \hfill \\
  Estimate total runtime up to \texttt{NP\_TOT} processors
\item[\texttt{--sf SER\_FRAC}] \hfill \\
  Fraction of serial code to estimate parallel runtimes (Default: 0.5)
\end{description}
If the ``\texttt{--t}'' runtime option is present, instead of running the set, \texttt{crv\_mp\_run\_set} prints out estimates of the cost of the calculation from 1 to \texttt{NP\_TOT} processors relative to the cost of a single edge calculations performed in serial mode.

\subsubsection{Auxiliary options}

\begin{description} \itemsep5pt \parskip0pt \parsep0pt
\item[\texttt{--v}] \hfill \\
  Generate verbose output (timings, etc)
\item[\texttt{--dr}] \hfill \\
  Dry run: Do everything but actually run
\end{description}

\subsubsection{Examples}

A typical instance of \texttt{crv\_mp\_run\_set} to run the set \texttt{IronOxides} created in the previous example (Fig. \ref{fig:feoout}) is as follows:

\texttt{crv\_mp\_run\_set --ppn 64 -nn 45 --v IronOxides}

\noindent or

\texttt{crv\_mp\_run\_set -np 2880 --v IronOxides}

\noindent Given that the set contains 45 edges, for maximum efficiency we assign one node with 64 processors to each edge, or a total of 2880 processors, thus effectively running all materials simultaneously. FEFF10 then computes each individual edge parallelizing over the energy grid with 64 processors.

\subsection{Requirements}

\subsubsection{Code distribution}
Both Corvus\cite{corvusgit} and FEFF10\cite{feffgit} are open source and are available on GitHub. Detailed installation instructions can be found in Ref. \cite{opconsinstall}.

\subsubsection{{\rm Corvus} requirements}

The Corvus workflow manager is implemented in Python 3. In addition, it requires the Numpy\cite{numpy} and Scipy\cite{scipy} packages, 
as well as a few other packages depending on the required functionality. 
The Corvus workflows that are possible will be determined by the specific external software packages installed on the system. 
In this case, the optical constants WF only requires the appropriate version of FEFF10.

\subsubsection{FEFF10 requirements}

To run efficiently, a parallel execution is important.  Thus, the FEFF code
requires a Fortran90 compiler with some flavor of MPI-2 for compilation.
FEFF has been tested with a wide variety of compilers and MPI-2 implementations.

\subsubsection{{\rm Corvus \texttt{[opcons]} } specific requirements}
\label{subsubsec:req}

The \texttt{crv\_mp\_mk\_set} and \texttt{crv\_mp\_run\_set} tools are also implemented in Python 3 and, 
in addition to the Corvus requirements, they need the pymatgen package.\cite{pymatgen} 
In order to use the MPRester method in pymatgen to access the Materials Project database directly, 
users need to have an MP account and a valid API key. For the work presented in this paper, we used
pymatgen v2022.0.9, and MP v2019.05.


\subsubsection {Performance Analysis}

The CPU time for a given material scales linearly with the number of edges and
unique atom positions in a unit cell and, depending on the SCF and FMS settings,
typically takes between a few minutes and $1/2$ hr. Thus, when the available
number of processors is equal or greater than the total number of edges, linear
parallelization is simply achieved by assigning a processor or more to each
edge, up to about 32-64 processors. More processors per edge becomes less
efficient at the level of FEFF calculations. In other cases, and for
high-throughput runs for many different materials, the wall-clock time depends
on how the set is structured (e.g., how many edges per material in the set) and
how the calculations are distributed in the workflow. In that case, there are no
simple rules to determine the optimal parallel distribution. To help with this
issue, the \texttt{crv\_mp\_run\_set} tool includes an option
(``\texttt{--t}'') to estimate how well a given system of interest will
parallelize for a certain range of number of processors (``\texttt{--np}'') relative to
the fully serial calculation. 



\begin{figure}[h]
\includegraphics[scale=0.32,clip,trim=0.0cm 0.0cm 0.0cm 0.0cm]{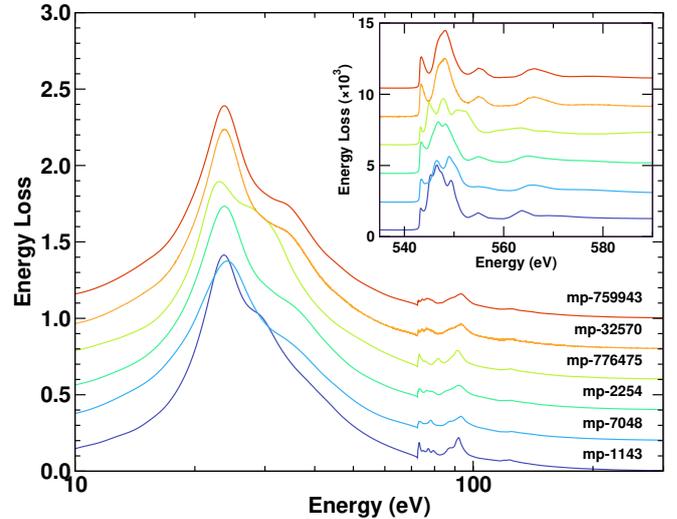}
\caption[Periodic Table of FEFF Optical Constants]
{Loss function for several Al$_2$O$_3$ oxides listed in the MP.
\label{fig:AlOxides}}
\end{figure}

\section{Example Applications}

\subsection{Optical constants of the elements}

As an illustration, we have carried out calculations for the most
stable structure (at room temperature and nominally 1 atm) of each of the elemental solids listed in the MP with data imported for a given MP-ID.  Some of the most important factors limiting the
quality of the results are the SCF and FMS radii used in the simulations. The results can be obtained online at \url{http://feff.phys.washington.edu/optical-constants/}\cite{opconsurl}  Fig.\ \ref{fig:rconv} (top) shows a comparison of the results using an increasing number of near-neighbor shells for $\epsilon_2$ of copper. We find that 3 and 6 shells in the SCF and FMS modules, respectively, produce converged results above ~10-20 eV.  

Overall, the results are in reasonable agreement with   experimental tabulations,\cite{desy} and illustrate   that, except close to an edge
or in the
optical region, $\epsilon_2$ can be approximated reasonably by the atomic approximation,
ignoring the fine structure.
For example, Fig.\ \ref{fig:rconv} shows a comparison to experiment\cite{desy} of Cu $\epsilon_2$ (top) and real index of refraction Au (bottom) computed with converged SCF and FMS radii and including the density of states convolution in Eq.\ \ref{eq:dosconv}.
These results show that Eq.\ \ref{eq:dosconv} is only semi-quantitative
in the visible range.
Fig.\ \ref{fig:ptable} shows a periodic table of the elements and associated MP-IDs that are currently available and can be freely viewed and
downloaded at \url{http://feff.phys.washington.edu/optical-constants/}.\cite{opconsurl}


\subsection{Optical constants of the Al$_2$O$_3$ materials in the MP}

In addition to the elemental solids, we have also computed several aluminum
oxides.
For example, Fig.\ \ref{fig:Al2O3loss} shows a comparison to
experiment\cite{desy} of the energy loss for the full energy range
for the corundum form of Al$_2$O$_3$.  In order to demonstrate the set
capabilities of the Corvus \texttt{[opcons]} workflow,
we calculated results for a sub-set
of those listed in the MP with formula Al$_2$O$_3$.
The results for the loss function up to the Al L-edges are shown in
the main panel of Fig.\ \ref{fig:AlOxides}, while the inset
highlights the O K-edge region. The differences in the near edge structure,
as well as those in 
the optical reflect the different symmetries present in the
system. For example, there is a mixture of Al sites in octahedral
and tetrahedral symmetry in the first two materials, which
are both monoclinic, while the last material is trigonal, and contains
only Al in octahedral symmetry. Similarly, the oxygen sites in the
first two systems are a mixture of sites in tetrahedral and trigonal
symmetry. These changes in symmetry lead to differences in electronic
structure, which are directly related to the features seen in the near
edge spectra.


\section{Conclusions}

We have presented an optical constants (``\texttt{opcons}'') workflow built with the Python-based 
Corvus workflow manager for high-throughput simulations of full spectrum optical
constants from the UV-VIS to hard x-ray wavelengths for materials defined in
the MP data-base. The calculations are based
on the efficient RSGF code FEFF10 and an approximate treatment of the valence
contributions using structural data from the MP database
specified by a unique MP materials ID. As for the XAS, the results can be 
represented in terms of an atomic-like background and an
optical fine-structure (OFS)
analogous to EXAFS. Except close to an edge where the fine-structure is
substantial, the background dominates and
can be calculated efficiently over the full spectrum. The fine-structure
is edge-specific and as in EXAFS, reflects the structure of the environment of
a given absorber in the material. The calculations are set up automatically
using \emph{Corvus} auxiliary tools, and the workflow parallelizes the FEFF
computations over edges and sites
with an optimal distribution of computational resources.
Representative results show that the workflow produces accurate
optical properties over a broad range of energies.
Although this algorithm includes mean-free paths and Debye-Waller factors,
multiplet splitting, excitonic effects, and long-range contributions to
the valence spectra are currently neglected.  Consequently, the present
results may only be semi-quantitative in the UV-VIS regime.

\section{Data availability}

We have computed the optical properties for all elemental solids in the periodic
table for which structures are available in the Materials Project database,
using the approach in this work. The data, including input files and structure,
are freely available online at
\url{http://feff.phys.washington.edu/optical-constants/}. Complete installation and
testing instructions as well as examples for the Corvus \texttt{[opcons]}
workflow, its associated tools, and the FEFF10 code are available at
\url{http://feff.phys.washington.edu/optical-constants/installation.html}.

\section*{Acknowledgments}

We thank C. Draxl, S. Dwaraknath, L. Hung, L. Reining,
and J. Vinson for comments and suggestions.
This work was supported primarily by the US Department of Energy, Office of
Basic Energy Sciences, Division of Materials Sciences and Engineering,
under Contract No.\ DE-AC02-76SF00515, specifically through the Theory
Institute for Materials and Energies Spectroscopies (TIMES)
program (JJR, JJK, FDV, and CDP), with computational support from NERSC, a
DOE Office of Science User Facility, under Contract No.\ DE-AC02-05CH11231.
R.~X.~Yang and K.~A.~Persson acknowledge support by the US Department of
Energy, Office of Science, Office of Basic Energy Sciences, Materials Sciences
and Engineering Division under Contract No.\ DE-AC02-05-CH11231 (Materials Project program KC23MP).


\bibliographystyle{elsarticle-num}
\biboptions{numbers,sort&compress}
\bibliography{mpopcons-vresub}

\end{document}